\documentclass[12pt,letterpaper]{article}
\usepackage[body={18cm, 25cm}]{geometry}
\usepackage[english, activeacute]{babel}
\usepackage[utf8]{inputenc}
\usepackage{lscape}
\usepackage{amsmath,amsfonts,amssymb}
\usepackage{graphicx}
\usepackage{geometry}
\usepackage[round]{natbib}
\usepackage{float}
\usepackage[margin=20pt,font=small,labelfont=bf,labelsep=period]{caption}
\usepackage{subcaption}  
\usepackage[pdftex, pdftitle={SFEM-phononic crystal}, pdfauthor={N. Guarin and J. Gomez}, pdfsubject={Numerical dispersion in FEM}, pdfkeywords={Numerical dispersion,Finite element method, Spectral element method, Wave propagation}, pdfpagemode=UseOutlines,bookmarks,bookmarksopen,pdfstartview=FitH,colorlinks,linkcolor=blue, urlcolor=black, citecolor=blue]{hyperref} 
\usepackage{cleveref}  

\begin{document}
\setlength{\parskip}{6mm}
\title{\textbf{Evaluation of the Spectral Finite Element Method With the Theory of Phononic Crystals.}}
\author{Nicol\'as Guar\'in-Zapata and Juan Gomez\\
        Departamento de Ingenier\'ia Civil\\
                Universidad EAFIT\\
                Medell\'in,\\
                Colombia\\
                \texttt{jgomezc1@eafit.edu.co}}

\maketitle
\begin{abstract}
We evaluated the performance of the classical and spectral finite element method in the simulation of elastodynamic problems. We used as a quality measure their ability to capture the actual dispersive behavior of the material. Four different materials are studied: a homogeneous non-dispersive material, a bilayer material, and composite materials consisting of an aluminum matrix and brass inclusions or voids. To obtain the dispersion properties, spatial periodicity is assumed so the analysis is conducted using Floquet-Bloch principles. The effects in the dispersion properties of the lumping process for the mass matrices resulting from the classical finite element method are also investigated, since that is a common practice when the problem is solved with explicit time marching schemes. At high frequencies the predictions with the spectral technique exactly match the analytical dispersion curves, while the classical method does not. This occurs even at the same computational demands. At low frequencies however, the results from both the classical (consistent or mass-lumped) and spectral finite element coincide with the analytically determined curves.
\end{abstract}


\section{Introduction}
During the recent years, and with particular interest steaming from the earthquake engineering community, the spectral finite element method (SFEM) has emerged as a powerful computational tool for numerical simulation of large scale wave propagation problems \citep{seriani95, faccioli97, komatitsch2004, magnoni2012spectral, regsem2012, shani2012simulation, Kudela2007, zhu2011, davies2009, Luo2009a,Luo2009b,Luo2009c,Balyan2012}. The origins of the currently used SFEM, can be traced back to the spectral techniques introduced by \cite{orszag1980spectral}, \cite{patera84} and \cite{Maday89}, initially proposed for fluid dynamics problems and \cite{gazdag1981modeling}, \cite{kosloff1982forward} and \cite{kosloff1990solution} in elastodynamics. In seismic wave propagation, the spectral element methods were first introduced by \cite{priolo1994numerical} and \cite{faccioli1996spectral}. Later \cite{komatitsch98} replaced the polynomials used by \cite{priolo1994numerical} by Legendre polynomials and incorporated a Gauss-Lobatto-Legendre quadrature, leading to the currently known version of the method in wave propagation problems. In these methods $\pi$-nodal points per wavelength are necessary to resolve the wave (compared to the 8-10 in the classical FEM) \citep{ainsworth2009}; they are also less sensitive to numerical anisotropy and element distortion; exhibit a smaller conditioning number and lead by construction, to diagonal mass matrices. The fundamental idea behind the SFEM is the use of higher order Lagrange interpolation at non-equidistant nodal points, reducing the so-called Runge-phenomenon and producing exponential convergence rates. If the nodal points, are also the quadrature points corresponding to a Gauss-Lobatto-Legendre integration rule, the resulting scheme yields diagonal mass matrices, whereby the global equilibrium equations from explicit time marching schemes become uncoupled. In contrast, one of the drawbacks of the spectral technique, is the fact that special meshing methodologies must be developed, since in the SFEM algorithm the nodal points must be placed at certain (non-standard) positions within the element. It is thus of interest to identify the range of frequencies for which the spectral technique results truly advantageous with respect to classical displacement based formulations. Although there are several works dealing with the performance of the spectral approach \citep{marfurt1984accuracy,dauksher2000solution,de2007grid,ainsworth2005,seriani2008dispersion,ainsworth2009,mazzieri2012dispersion}, these have been mainly restrained to problems with particular boundary conditions or else, have been conducted using strongly simplified assumptions.

In this article we evaluate the capabilities of the SFEM in the simulation of mechanical wave propagation problems. We compare its performance with the classical finite element algorithm using as a quality index, numerically obtained dispersion curves of the propagation material, computed with the aid of the theory of phononic crystals. This is an objective performance metric, since it qualifies the ability of the method to describe the material and not a particular wave propagation problem, thus the effects of geometry and boundary conditions appearing in the solution of a finite model are not taken into account in this model. For purposes of the evaluation, we implemented classical and spectral finite element algorithms and found numerical dispersion curves for four different materials, namely a homogeneous material cell, a bilayer material cell, a porous material cell and a composite material cell. We contrasted the numerical curves obtained with both methods, with analytic solutions and identified non-dimensional frequency ranges, where the classical and spectral finite element methods deviated from each other and from the analytic solution. We also evaluated the error, introduced in the dispersion properties of the material by the artificial diagonalization process of the finite element mass matrix. This practice is commonly encountered in explicit time marching schemes implemented with the classical algorithm. From our study we found that the spectral finite element method yields reliable results at much higher frequencies than the classical technique even at the same computational costs. At low frequencies however both methods give the same precision which clearly favours the classical treatment.

In the first part of the article we briefly describe the implementation of the spectral finite element method. We then discuss general aspects regarding the numerical formulation of the eigenvalue problem corresponding to the imposition of the Floquet-Bloch boundary conditions for a spatially periodic material. In the final part of the article we present results, in terms of dispersion curves for different material cells all of which are solved numerically and analytically.

\section{The Spectral Element Method}
\begin{figure}[H]
\centering
\includegraphics[height=7cm]{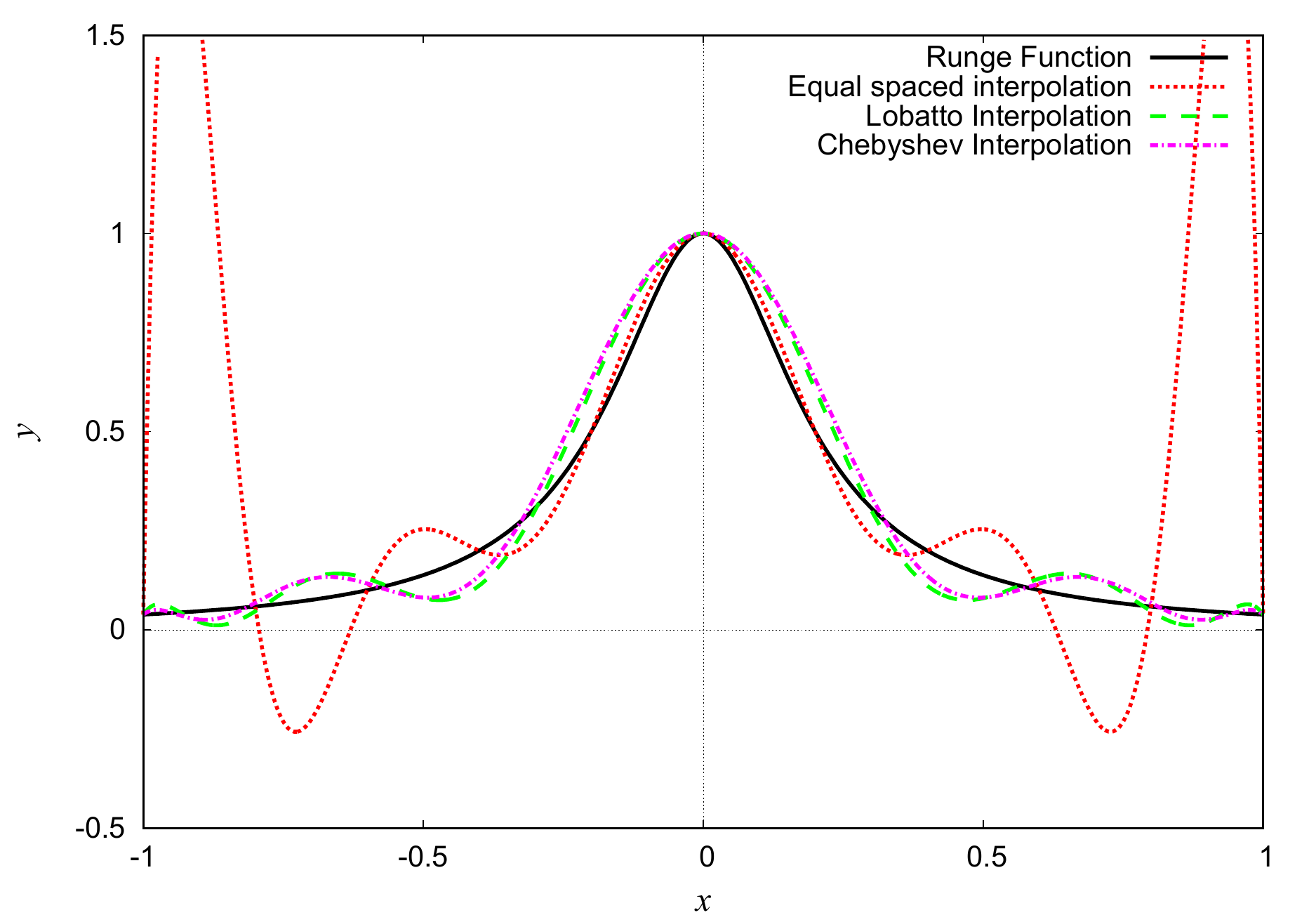} 
\caption{Lagrange interpolating polynomials of order 10 for the Runge function $\frac{1}{1+25x^2}$.}
\label{fig:runge_interp}
\end{figure}

The Spectral Finite Element Method (SFEM) is a finite element technique based on high order Lagrange interpolation \citep{book:zienkiewicz_FEM2} but with a non-equidistant nodal distribution. The above difference in the spatial sampling, with respect to classical algorithms, seeks to minimize the Runge phenomenon \citep{book:pozrikidis}. This corresponds to the spurious oscillations appearing at the extremes of an interpolation interval when--in an attempt to perform $p$-refinement--the number of sampling points is increased while keeping an equidistant nodal distribution. The problem is illustrated in Figure \ref{fig:runge_interp}, where a Runge function is interpolated with  3 different Lagrange polynomials of order 10, but with a different distribution of the sampling points. We used equidistant nodal points (as in the classical FEM algorithm) and also Lobatto and Chebyshev nodes (as in the SFEM algorithm).

The Runge phenomenon corresponds to the strong oscillations observed towards the ends of the interval associated with the equidistant nodal distribution. In contrast, when the interpolation interval is sampled at the Lobatto and Chebyshev nodes, the approximation describes closely the Runge function throughout the interval. Since the method uses high-order interpolation for each element, it in fact corresponds to a $p$-refinement version of the classical technique, but with a different spatial sampling inside the elements. The term \emph{spectral} refers to the rate of convergence since the method converges faster than any power of $1/p$, where $p$ is the order of the polynomial expansion \citep{book:pozrikidis}.
There are two nodal distributions commonly used in spectral finite element methods: Chebyshev nodes and Lobatto nodes. The Chebyshev nodes are the roots of the Chebyshev polynomials of the first kind. The resulting interpolation polynomial not only  minimizes the Runge phenomenon but it also provides the best approximating polynomial under the maximum norm \citep{book:burden2011,book:kreyzsig_functional}. The Chebyshev polynomials of the first kind $T_n(x)$ are defined as the unique polynomial satisfying
\[T_n(x) = \cos\left[ n \arccos x\right]\quad \mbox{for each } n\geq 0,\mbox{ and } x\in [-1,1] \enspace ,\]
and with roots corresponding to
\[x_k = \cos\left( \frac{2k-1}{2n} \pi\right) \enspace .\]
 
\begin{figure}[h]
\centering
	\begin{subfigure}[t]{0.2\textwidth}
		\includegraphics[width=\textwidth]{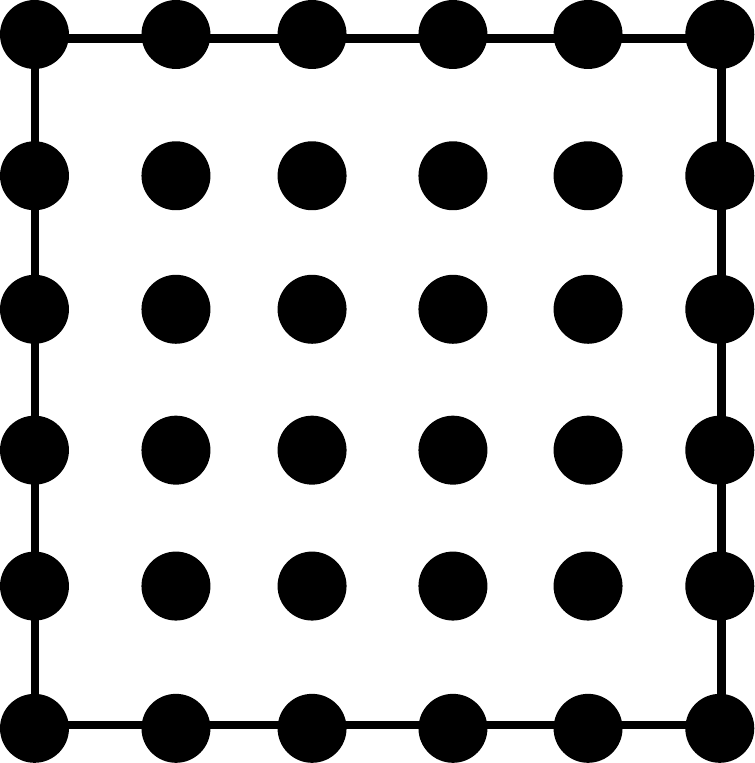}
		\caption{Classical $6\times 6$ element.}
	\end{subfigure}\qquad
	\begin{subfigure}[t]{0.2\textwidth}
		\includegraphics[width=\textwidth]{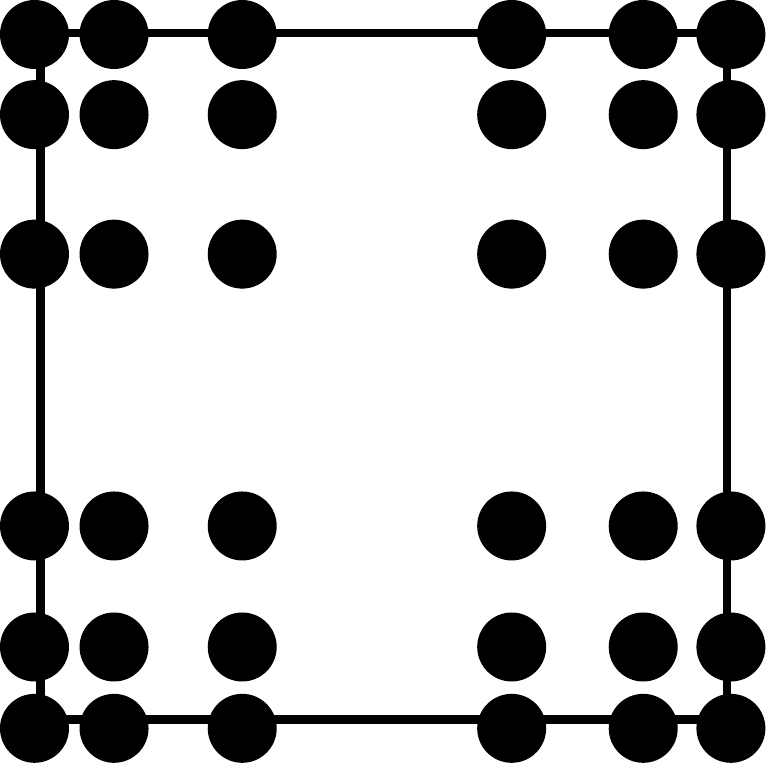}
		\caption{Lobatto-spectral $6\times 6$ element.}
	\end{subfigure}
\caption{Comparison of a classical and a Lobatto-spectral element.}
\label{fig:els_comparison}
\end{figure}

Similarly, the Lobatto nodes are the roots of the Lobatto polynomials, which are the first derivative of the Legendre polynomials
\[Lo_{i-1}(x) = P'_{i} (x)\quad x\in [-1,1]\enspace ,\]
with
\[P_{i}(x) = \frac{1}{2^i i!} \frac{d^{(i)}(t^2 -1)^i}{dt^{(i)}} \enspace ,\]
being the $i$th Legendre Polynomial \citep{book:abramowitz65}. The Lobatto nodes are also the Fekete points for the line, the square and the cube. The Fekete points in an interpolation scheme correspond to the nodal distribution that maximizes the determinant of the Vandermonde matrix \citep{bos2001, book:pozrikidis}. If the nodes (corresponding to the sampling points in the interpolation scheme) are also used as quadrature points in the computation of volume and surface integrals arising in the FEM, diagonal mass matrices are obtained due to the discrete orthogonality condition \citep{book:pozrikidis}. This last feature of the SFEM makes it very attractive in explicit time marching schemes, since this condition uncouples the equilibrium equations. In the classical displacement-based FEM this lumping process is artificially enforced, while in the SFEM it is a natural result. The nodal distributions corresponding to the $6\times6$ classical and Lobatto-spectral elements are shown in Figure \ref{fig:els_comparison}.

In this work we use the SFEM to solve the reduced elastodynamic frequency domain equation. The formulation follows from the minimization of the total potential energy functional $\Pi=\Pi(u_i)$ \citep{book:reddy_functional_analysis} leading to the principle of virtual displacements at the frequency $\omega$ given by
\begin{equation}
\int\limits_{\Omega} \sigma^{*}_{rs}(\mathbf{x}) \delta\epsilon_{rs}(\mathbf{x}) d\Omega -\omega^2 \int\limits_{\Omega} \rho(\mathbf{x}) u_r^{*}(\mathbf{x}) \delta u_r(\mathbf{x})d\Omega - \int\limits_{\Gamma} t_r^{*}(\mathbf{x}) \delta u_r(\mathbf{x}) d\Gamma - 
\int\limits_{\Omega} f_r^{*}(\mathbf{x}) \delta u_r(\mathbf{x}) d\Omega =0
\label{eq:weak_form}
\end{equation}
and where $\Omega=$ solution domain, $\Gamma=$ boundary, $\sigma_{rs}=$ stress tensor, $t_{r}=$ tractions vector, $\epsilon_{rs}=$ strain tensor, $u_r=$ displacement vector, $f_r=$ body force vector and $\mathbf{x}=$ field point vector. The superscripts $*$ appearing in \eqref{eq:weak_form} refer to complex conjugate variables. This is a requirement in the weak form, that produces a consistent inner product between complex variables and leads to self-adjoint operators and therefore Hermitian finite element matrices: this last condition guarantees the existence of real eigenvalues. The discrete finite element equations follow after writing in \eqref{eq:weak_form} the displacements (and related variables) in terms of basis functions or its derivatives as
\[\mathbf{u}_i(\mathbf{x})=N_i^{Q} (\mathbf{x}) \mathbf{U}^Q\]
where $\mathbf{u}_i(\mathbf{x})$ is the displacement vector evaluated at point $\mathbf{x}$, $N_i^Q(\mathbf{x})$ is the shape function for the $Q$th node and $\mathbf{U}^Q$ is the nodal displacement vector at the $Q$th node. In our notation we retain the index $i$ in the shape function just to indicate that a vectorial variable is being interpolated. The final finite element equations are of the familiar form
\begin{equation}
[K-\omega^2M] \left\lbrace \mathbf{U} \right\rbrace - \left\lbrace \mathbf{F} \right\rbrace = \left\lbrace \mathbf{0} \right\rbrace \enspace .
\label{eq:discrete_form}
\end{equation}
where $[K]$ and $[M]$ are complex valued Hermitian stiffness and mass matrices and $\mathbf{F}$ is the nodal forces vector comprising boundary tractions and body forces. In 2D the Shape functions for the $Q-th$ node can be computed as the product of independent Lagrange polynomials like
\begin{equation}
N_i^Q(x,y) = \ell^{Q_x} (x) \ell^{Q_y}(y) \enspace .
\end{equation}
Similarly, the derivatives of the shape functions are
\begin{equation}
\frac{\partial N_i^Q(x,y)}{\partial x} = \ell'^{Q_x}(x) \ell^{Q_y}(y),\quad 
\frac{\partial N_i(x,y)}{\partial y} = \ell^{Q_x}(x) \ell'^{Q_y}(y) \enspace ,
\label{eq:interp}
\end{equation}
with 
\begin{align}
\ell^Q(x) &= \prod\limits_{{\substack{P=1 \\ Q\neq P }}}^{M+1} \frac{x - x^P}{x^Q - x^P} \enspace, \\
\ell'^Q(x) &= \sum\limits_{{\substack{P=1 \\ Q\neq P }}}^{M+1} \frac{1}{x - x^P}
\prod\limits_{\substack{R=1 \\ R\neq P }}^{M+1} \frac{x - x^R}{x^Q - x^R} \enspace ,
\end{align}
where $M$ is the order of the Lagrange polynomial.

If the integrals in \eqref{eq:weak_form} are numerically integrated using the Gauss-Lobatto nodes, which at the same time are the element nodal points, a diagonal mass matrix is obtained. Although this is advantageous in the case of an explicit time marching scheme, only polynomials of order $2N-3$ \citep{book:abramowitz65,book:pozrikidis} or smaller are integrated exactly. This accuracy should be contrasted with the one in the standard Gauss-Legendre quadrature, that integrates exactly polynomials of order $2N-1$ or smaller \citep{book:abramowitz65,book:bathe}. This reduction in the integration accuracy is due to the inclusion of nodes in the extremes of the interval. Since the order of exact integration decreases in the Gauss-Lobatto case, that selection is useful for interpolations with order higher than $N=3$, because the nodal distribution in both methods is the same for $N=1,2$. For the Gauss-Lobatto quadrature the nodes are the zeros of the completed Lobatto polynomials of order $N$
\[(1-x^2) L'_N(x)=0\]
with weights
\begin{align*}
w_1 &=w_{N+1} = \frac{2}{N(N+1)}\\
w_i &= \frac{2}{N(N+1)}\frac{1}{L_N^2(x_i)}\quad \mbox{for } i=2,3,\cdots,N \enspace .
\end{align*}

\section{Phononic Crystals}
Phononic crystals may be synthetic or natural materials, formed by the spatially periodic repetition of a microstructural feature. Such materials consist of a number of identical structural components joined together in an identical fashion into an elementary cell that repeats periodically throughout space, forming the complete material and described by a lattice construction \citep{mead1996wave}. When subjected to waves propagating at high enough frequencies, the microstructures behave as micro-scatterers, translating into dispersive propagation behaviour at the macroscopic level and sometimes even leading to states of imaginary group velocity, i.e., values of frequency for which waves do not propagate denoted as bandgaps. At even higher frequencies, additional propagation modes may be triggered once a certain cut-off value of the frequency is reached. The relation between the frequencies at which propagation is possible and the material periodic microstructure--described in terms of the wave vector--can be elegantly described in terms of the dispersion curve.

In the last few years a strong effort has been devoted to the study of propagation of waves in periodic materials. In particular, methods to determine--via numerical techniques--the dispersion curves for a specific material, have emerged from earlier work in the field of solid-state physics, \cite{book:brillouin}. The dispersion curve can be obtained through the analysis of a single cell, after having identified the microstructure that repeats periodically. Within that context, the unit cell is described by lattice base vectors indicating the directions that must be followed, in order to cover the complete space with the application of successive translation operations. As described in \cite{Srikantha2006}, the joints of any lattice structure can be envisioned as a collection of points (lattice points). These, at the same time are associated with a set of basis vectors $\mathbf{a}_i$. The entire direct lattice is then obtained after tessellating the elemental cell along the basis vectors $\mathbf{a}_i$. In this way the location of any point in the periodic material can be identified with $n_i$ integers, defining the number of translations of the elemental cell along the $\mathbf{a}_i$ directions of the basis vectors.

With the aid of the fundamental tool provided by Bloch's theorem \citep{book:brillouin} and extracted from the theory of solid-state physics, the dispersive properties of the material can be obtained via finite element analysis of the elementary cell. That theorem states that the change in wave amplitude occurring from cell to cell does not depend on the wave location within the periodic system \citep{ruzzene2003}. The final statement corresponding to Bloch's theorem, is written in terms of phase shifts between displacements and tractions on opposite surfaces of the elemental cell like
\begin{equation}
\begin{split}
\mathbf{u_i}(\mathbf{x} + \mathbf{a}) = \mathbf{u_i}(\mathbf{x}) e^{i \mathbf{k} \cdot \mathbf{a}} \enspace ,\\
\mathbf{t_i}(\mathbf{x} + \mathbf{a}) = -\mathbf{t_i}(\mathbf{x}) e^{i \mathbf{k} \cdot \mathbf{a}} \enspace
\end{split}
\label{eq:bloch_theorem}
\end{equation}
and where $\mathbf{k}$ is the wave vector and $\mathbf{a} = \mathbf{a}_1 n_1 + \mathbf{a}_2 n_2 + \mathbf{a}_3 n_3$ is the lattice translation vector, formed after scaling the basis vectors by the translation integers $n_i$. The factor $e^{i \mathbf{k} \cdot \mathbf{a}}$ corresponds to the phase shift between opposite sides of the cell. The real and imaginary components of the wave vector differ by this factor, while the magnitude remains the same. 

\subsection{Bloch-periodicity in discrete methods}
The formulation of the finite element equations for the determination of the dispersion properties of an elemental cell requires the assembly of the discrete equations for the complete cell, the subsequent reduction of the equations after imposition of the Bloch-periodic boundary conditions stated in \eqref{eq:bloch_theorem}, and the solution of a series of generalized eigenvalue problems to find the dispersion curve. If the eigenvalue problem or numerical dispersion relation is written like
\begin{equation}
\hat{D}(\mathbf{k},\omega)\mathbf{U}=0
\label{eq:dispersion}
\end{equation} 
where  $\mathbf{k}$ is the wave vector which is progressively assigned successive values, in such a way that the first Brillouin zone is fully covered. Each application of the wave vector and the solution of the related eigenvalue problem yields a duple $(\mathbf{k},\omega)$ representing a plane wave propagating at frequency $\omega$.

\begin{figure}[h]
\centering
	\begin{subfigure}[b]{0.3\textwidth}
		\includegraphics[width=\textwidth]{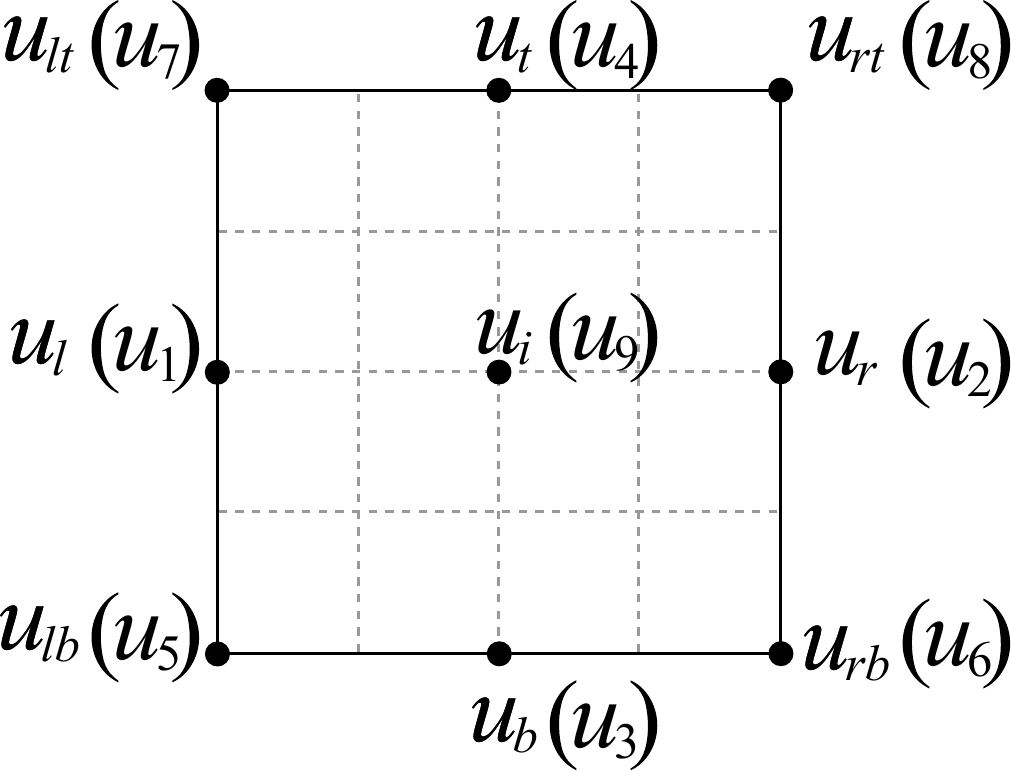}
		\caption{Original cell.}
	\end{subfigure}\quad
	\begin{subfigure}[b]{0.3\textwidth}\qquad
		\includegraphics[width=\textwidth]{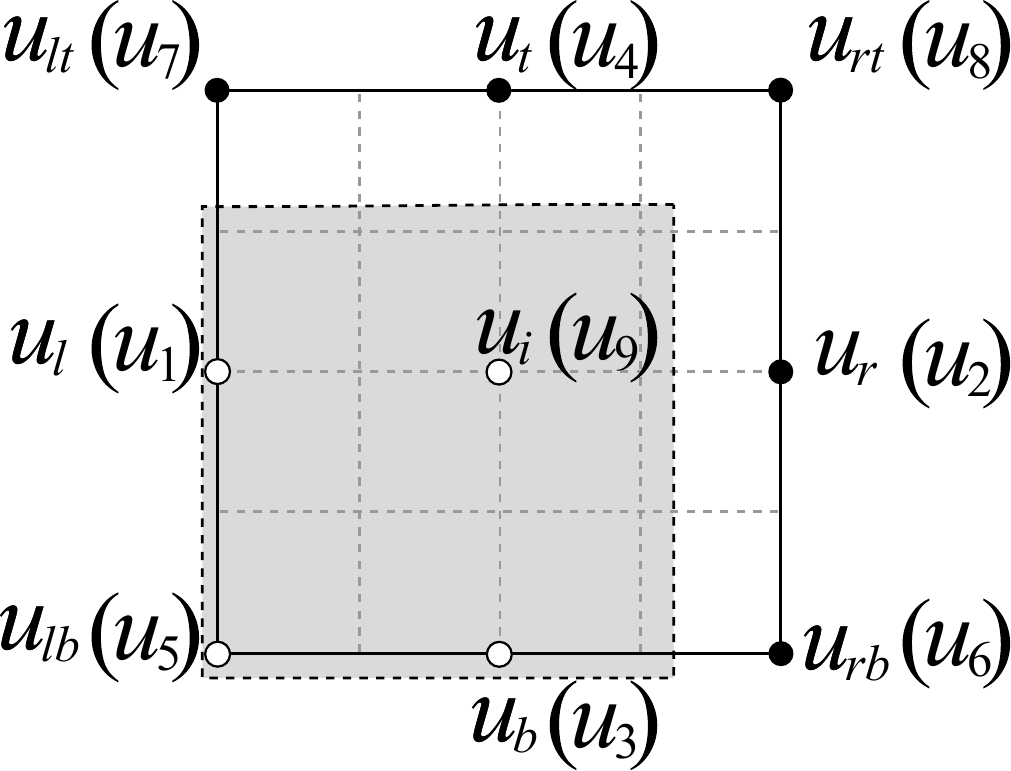}
		\caption{\emph{Reduced} cell.}
	\end{subfigure}\\
	\begin{subfigure}[b]{0.7\textwidth}\qquad
		\centering
		\includegraphics[width=0.45\textwidth]{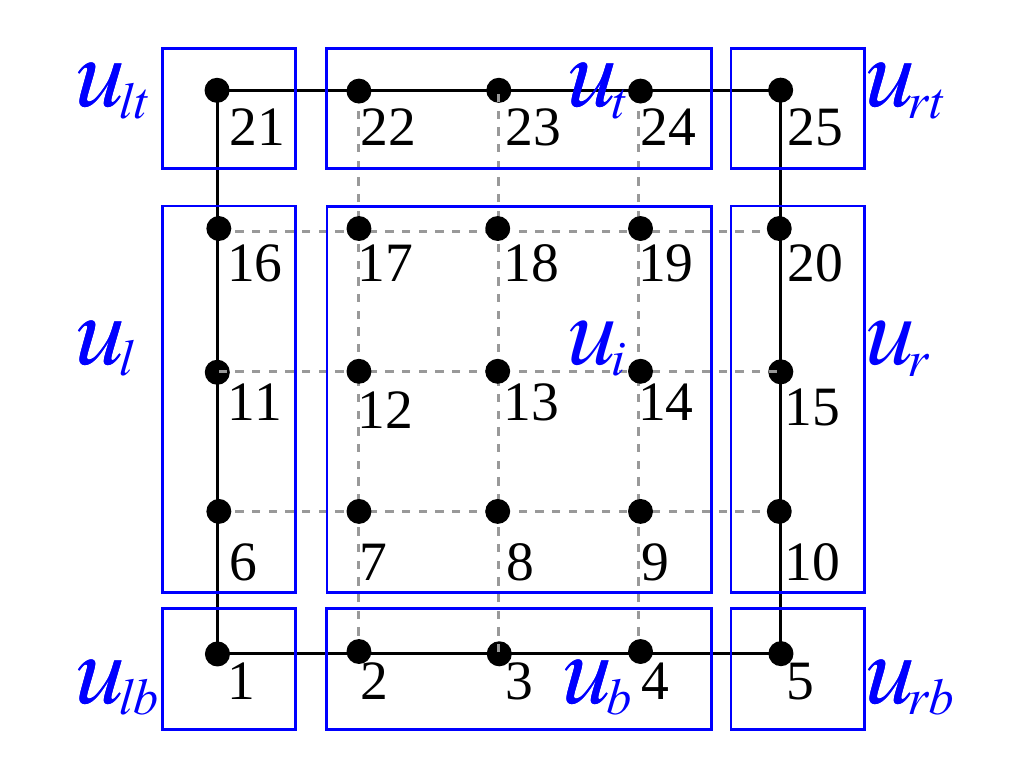}
		\caption{The black circles now correspond to single degrees of freedom. Notice that the nodal numbering does not match the original numbering for groups of degrees of freedom.}
	\end{subfigure}
\caption{Relevant sets of DOF for an schematic unitary cell discretized with several finite elements. Each circle represents a family of degrees of freedom for a typical mesh and not a single degree of freedom. In part a) we show all the degrees of freedom for the unitary cell before imposing the relevant Bloch-periodic boundary conditions. Part b) shows the reduced cell where the white circles enclosed by the dark square represent the \emph{reference} nodes containing the information from the \emph{image} nodes which will be eventually deleted from the system. Part c) shows an example of node grouping for a mesh of $4\times 4$ elements.}
\label{fig:bloch_FEM}
\end{figure}

The basic steps are further elaborated with reference to Figure \ref{fig:bloch_FEM}, describing a unit cell and where $2d_a$ and $2d_b$ are respectively the cell width and cell height. The sub-indices labelled as $9$ (or $i$) are related to the internal degrees of freedom, while those labelled $1-4$ ( or alternatively $l, r, b, t)$ are used for degrees of freedom over the boundary and $5-8$ (or $lb, rb, lt, rt$) correspond to those over the corners of the unit cell \citep{Srikantha2006}. In the initial step we assembly the complete discrete finite element equations for the unit cell using the following block vectors of nodal displacements and forces
\begin{align*}
\mathbf{U} &= [\mathbf{u}_l\ \mathbf{u}_r\ \mathbf{u}_b\ \mathbf{u}_t\ \mathbf{u}_{lb}\ \mathbf{u}_{rb}\ \mathbf{u}_{lt}\ \mathbf{u}_{rt}\ \mathbf{u}_{i}]^T \equiv [ \mathbf{u}_1\ \mathbf{u}_2\ \mathbf{u}_3\ \mathbf{u}_4\ \mathbf{u}_5\ \mathbf{u}_6\ \mathbf{u}_7\ \mathbf{u}_8\ \mathbf{u}_9]^T\ , \\
\mathbf{F} &= [\mathbf{f}_l\ \mathbf{f}_r\ \mathbf{f}_b\ \mathbf{f}_t\ \mathbf{f}_{lb}\ \mathbf{f}_{rb}\ \mathbf{f}_{lt}\ \mathbf{f}_{rt}\ \mathbf{f}_{i}]^T \equiv [ \mathbf{f}_1\ \mathbf{f}_2\ \mathbf{f}_3\ \mathbf{f}_4\ \mathbf{f}_5\ \mathbf{f}_6\ \mathbf{f}_7\ \mathbf{f}_8\ \mathbf{f}_9]^T\ .
\end{align*}

In the FEM equations the Bloch conditions take the form of constraint relationships among the degrees of freedom in opposite boundaries of the cell.  This allows reduction of the system of equations via elimination of the subset of image nodes and leaving only the subset of reference nodes. These constraints are written like
\begin{align*}
\mathbf{u}_2 &= e^{i \psi_x}\mathbf{u}_1\, \\
\mathbf{u}_4 &= e^{i \psi_y}\mathbf{u}_3\, \\
\mathbf{u}_6 &= e^{i \psi_x}\mathbf{u}_5\, \\
\mathbf{u}_7 &= e^{i \psi_y}\mathbf{u}_5\, \\
\mathbf{u}_8 &= e^{i (\psi_x + \psi_y)}\mathbf{u}_5,
\end{align*}
where $\psi_x = 2 k_x d_a $ and $\psi_y = 2 k_y d_b $ define phase shifts in $x$ and $y$ directions respectively and $(k_x,k_y)=\mathbf{k}$ corresponds to the wave vector. Similarly, using the corresponding relationships between the nodal forces we have
\begin{align*}
&\mathbf{f}_2 + e^{i \psi_x}\mathbf{f}_1=\mathbf{0}\, \\
&\mathbf{f}_4 + e^{i \psi_y}\mathbf{f}_3=\mathbf{0}\, \\
&\mathbf{f}_8 + e^{i \psi_x}\mathbf{f}_6 + e^{i \psi_y}\mathbf{f}_7 +e^{i (\psi_x + \psi_y)}\mathbf{f}_5= \mathbf{0} .\enspace
\end{align*}
If we assume that the problem does not involve internal forces, i.e. $\mathbf{f}_9=\mathbf{0}$, we get the reduced generalized eigenvalue problem to be solved for each frequency and related propagation mode and given by
\begin{equation}
[K_R-\omega^2 M_R]\lbrace \mathbf{U}_R\rbrace = \lbrace \mathbf{0}\rbrace \enspace .
\label{eq:reduced}
\end{equation}
In a finite element implementation the Bloch-periodic boundary conditions can be considered either including directly the phase shifts in the basis functions at the outset or performing row and column operations in the assembled matrices, \cite{sukumar_bloch-2009}. The Hermitian matrices $K_R$ and $M_R$ in the reduced system are function of the wave vector and angular frequency as indicated in \eqref{eq:dispersion}.

\section{Results}
In order to test the performance of the spectral finite element method, we implemented an in-house code to assemble and solve the generalized eigenvalue problem stated in \eqref{eq:reduced}. We considered a homogeneous (non-dispersive material), a bilayer material and a composite material made out of an aluminium matrix with a square inclusion. In particular we studied the case of a pore and a brass inclusion. From the above set, the homogeneous and bilayer material cells had closed form solutions, so we were able to compare our numerical results with the analytical dispersion curves.

In all the considered cases we solved the problem using both the classical finite element technique and the spectral finite element method. On the other hand, and since one of the reported advantages of the spectral technique, with respect to the classical approach, is the fact that in this method a diagonal mass matrix is obtained, we also tested the effect of using artificially lumped mass matrices in the classical FEM. With this goal in mind, we also solved the homogeneous cell and the bilayer material cell, using the classical FEM with a consistent mass matrix and a lumped mass matrix. The method for making the matrices diagonals is the \emph{Diagonal scaling} (see \cite{book:zienkiewicz_FEM1}), where each of the terms over the diagonal are scaled by a factor under the constraint of keep the total mass of the element constant. In our case, the factor is the total mass of the element ($M_{\text{tot}}$) divided by the trace of the matrix ($\mathrm{Tr}(M)$), i.e.
\[M^{(\text{lumped})}_{ii} = \frac{M_{\text{tot}}}{\mathrm{Tr}(M)} M_{ii} \quad \text{(no summation on $i$)} \enspace .\]

The results are presented next in terms of dimensionless frequency vs dimensionless wave number.

\subsection{Homogeneous material}
In an ideal homogeneous material a plane propagates non-dispersively according to the linear relation
\begin{equation}
\omega = c \Vert \mathbf{k} \Vert.
\label{eq:homogeneous}
\end{equation}
If the wave is travelling parallel to the $y$-axis (i.e., vertical incidence) and we consider a square cell of side $2d$ the above relation particularizes to 
\begin{equation}
\omega_i = c_i\sqrt{\left(k+\frac{n\pi}{d}\right)^2 + \left(\frac{m\pi}{d}\right)^2} \enspace ,
\label{eq:homog}
\end{equation}
where $c_i$ is the wave propagation velocity for the $P$ wave $(i=1)$ or $SV$ wave $(i=2)$. In this relation $n$ and $m$ are integers indicating the relative position of any given cell with respect the reference cell. When $n$ is different from zero, the dispersion branches become hyperbolas due to the wave vector whose components are determined as modulus $2\pi$ and could be present in any Brillouin zone. These hyperbolas result from the intersection of a plane parallel to the $x-$axis and a cone, \cite{Langlet1993, Langlet1995, MSc_thesis-Guarin2012}.

\begin{figure}[H]
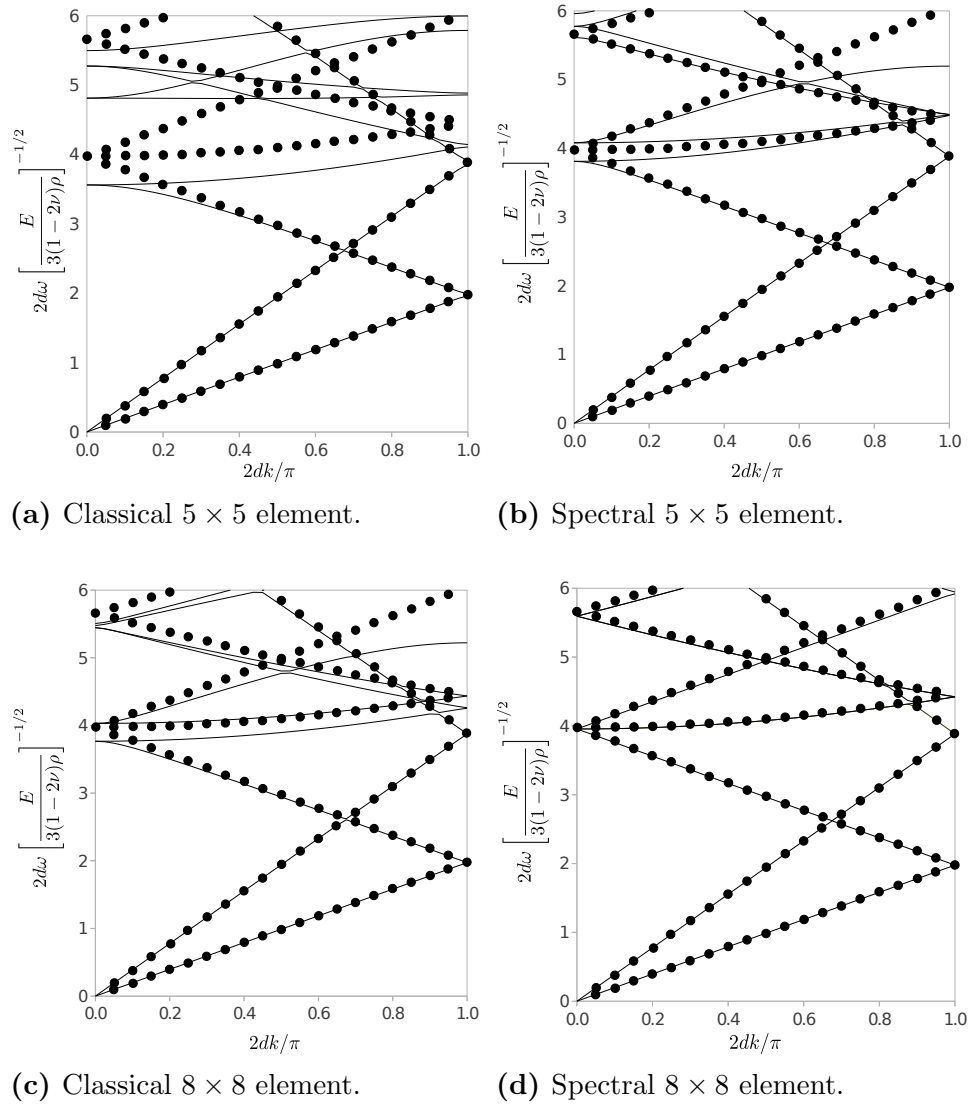

\centering
	\begin{subfigure}[b]{0.4\textwidth}\qquad
		\includegraphics[width=\textwidth]{homog-5x5-FEM.pdf}
		\caption{Classical $5\times 5$ element.}
	\end{subfigure}\,
	\begin{subfigure}[b]{0.4\textwidth}\qquad
		\includegraphics[width=\textwidth]{homog-5x5-SEM.pdf}
		\caption{Spectral $5\times 5$ element.}
	\end{subfigure}\\
	\begin{subfigure}[b]{0.4\textwidth}\qquad
		\includegraphics[width=\textwidth]{homog-8x8-FEM.pdf}
		\caption{Classical $8\times 8$ element.}
	\end{subfigure}\,
	\begin{subfigure}[b]{0.4\textwidth}\qquad
		\includegraphics[width=\textwidth]{homog-8x8-SEM.pdf}
		\caption{Spectral $8\times 8$ element.}
	\end{subfigure}
\caption{Dispersion curves from a single homogeneous material cell.}
\label{fig:homog-a}
\end{figure}

Figure \ref{fig:homog-a} presents the results for a square cell discretized with 4-th and 7-th order complete Lagrange polynomials with equidistant and Lobatto sampling nodes. The analytical and numerical dispersion curves are shown in dotted and continuous lines respectively. In all the considered cases the differences between the numerical and analytical results are due to numerical dispersion. The results from the classical 5$\times$5 element are accurate only at normalized frequencies below 3.0, while the analogous analysis with the spectral method predicts even modes of higher multiplicity as seen from the point near (1,4). Moreover, the classical 8$\times$8 element only captures the first three modes, while this same result was already reached with the 5$\times$5 spectral element. This of course is a result of the improved convergence properties of the spectral technique. Similarly the 8$\times$8 spectral element captures the solution exactly up to the sixth mode. 

\subsection{Bilayer material}
\begin{figure}[h]
\centering
\includegraphics[height=4cm]{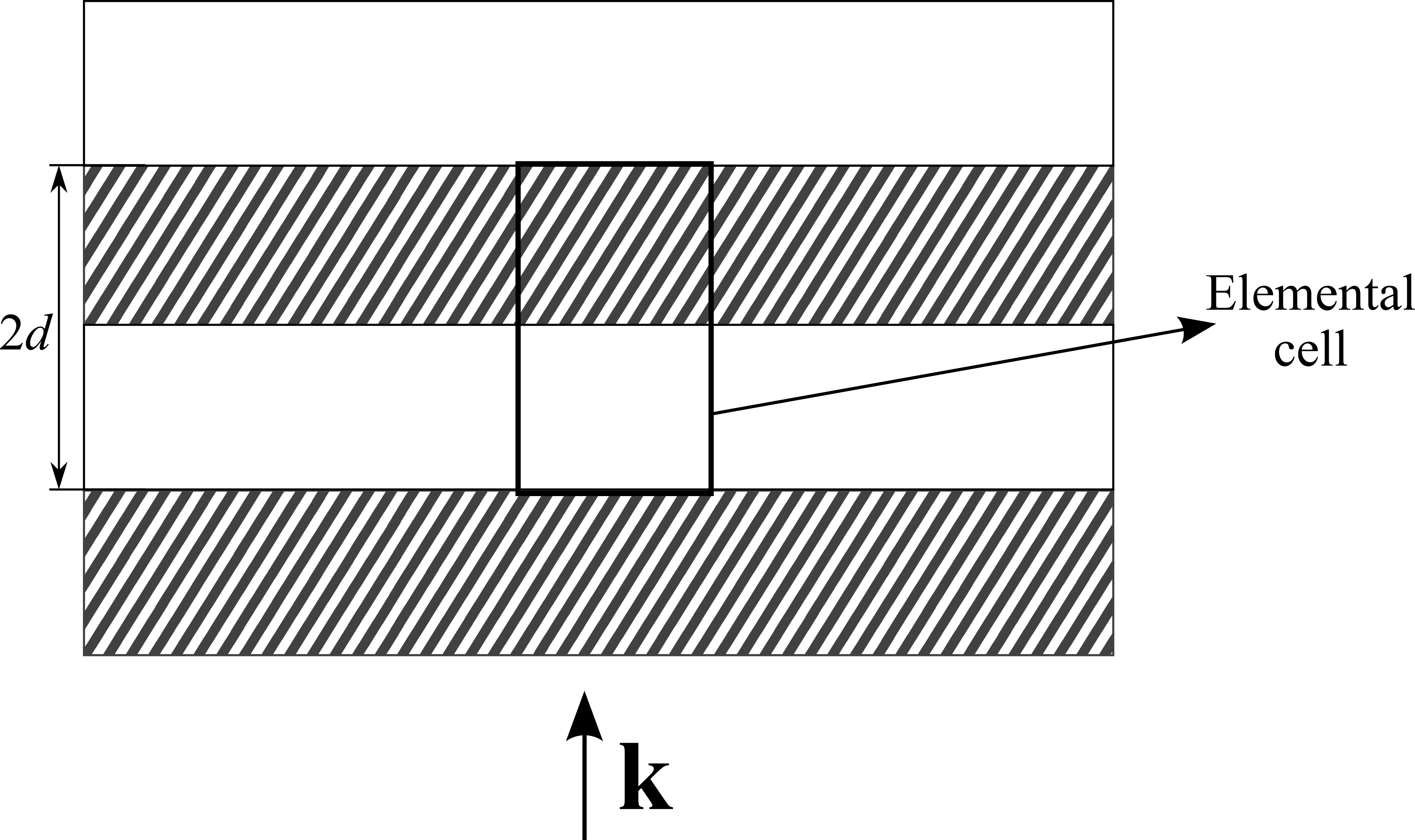}
\caption{Cell for a bilayer material.}
\label{fig:bilayercell}
\end{figure}

\begin{figure}[H]
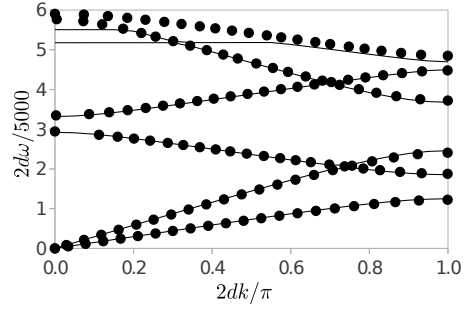
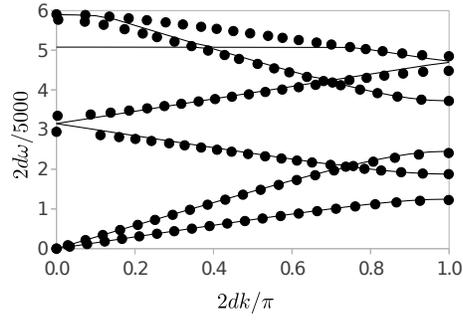
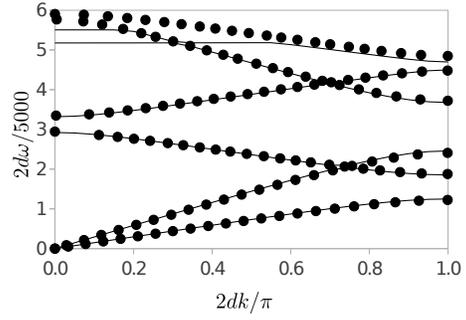
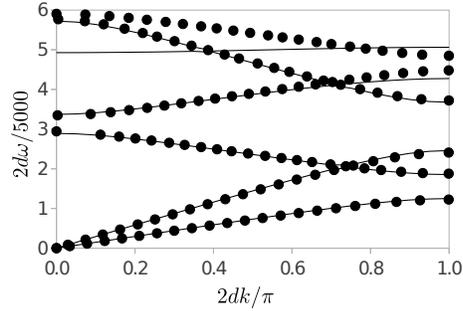
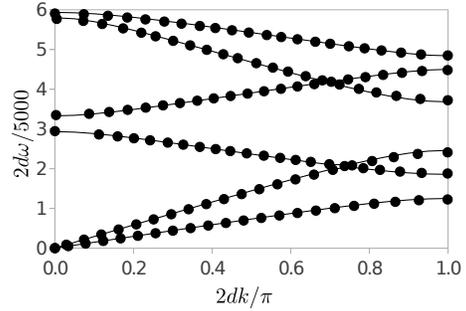

\centering
	\begin{subfigure}[b]{0.4\textwidth}\qquad
		\includegraphics[width=\textwidth]{bilayer-4x4-FEM.pdf}
		\caption{Classical $4\times 4$ element.}
	\end{subfigure}\,
	\begin{subfigure}[b]{0.4\textwidth}\qquad
		\includegraphics[width=\textwidth]{bilayer-8x8-FEM.pdf}
		\caption{Classical $8\times 8$ element.}
	\end{subfigure}\\
	\begin{subfigure}[b]{0.4\textwidth}\qquad
		\includegraphics[width=\textwidth]{bilayer-4x4-lumped_FEM.pdf}
		\caption{Classical $4\times 4$ lumped element.}
	\end{subfigure}\,
	\begin{subfigure}[b]{0.4\textwidth}\qquad
		\includegraphics[width=\textwidth]{bilayer-8x8-lumped_FEM.pdf}
		\caption{Classical $8\times 8$ lumped element.}
	\end{subfigure}\\
	\begin{subfigure}[b]{0.4\textwidth}\qquad
		\includegraphics[width=\textwidth]{bilayer-4x4-SEM.pdf}
		\caption{Spectral  $4\times 4$ element.}
	\end{subfigure}\,
	\begin{subfigure}[b]{0.4\textwidth}\qquad
		\includegraphics[width=\textwidth]{bilayer-8x8-SEM.pdf}
		\caption{Spectral  $8\times 8$ element.}
	\end{subfigure}
\caption{Dispersion curves for a bilayer material. Each layer is meshed with a single element with homogeneous material, being the materials aluminium and brass.}
\label{fig:bilayer_sol}
\end{figure}

\begin{figure}[H]
\centering
\includegraphics[width=0.45\textwidth]{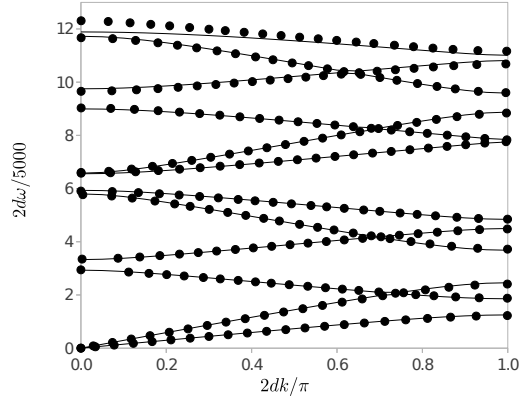} 
\caption{Results for 12 branches of the dispersion relation for the bilayer material computed with a single $8\times 8$  SFEM element for layer. The materials are aluminium and brass.}
\label{fig:bilayer_sol2}
\end{figure}

Figure \ref{fig:bilayercell} describes a composite material formed by two layers of different properties. When the material cell is submitted to an incident wave in the direction perpendicular to the layers the dispersion curve can be found in closed-form. The problem was studied by Lord Rayleigh in the case of electromagnetic waves which, in this case, is mathematically equivalent to the mechanical problem \citep{rayleigh1888}.
Of particular interest in this material is the presence of band gaps in addition of its dispersive behaviour. The analytic dispersion relation as obtained by \cite{rayleigh1888} is given by
\[
\cos(2dk) = \cos\left(\frac{\omega d}{c_1}\right)\cos\left(\frac{\omega d}{c_2}\right) - \frac{(\rho_1 c_1)^2 + (\rho_2 c_2)^2}{2\rho_1\rho_2 c_1 c_2}\sin\left(\frac{\omega d}{c_1} \right)\sin\left( \frac{\omega d}{c_2}\right) \enspace ,
\]
where $c_i$ the longitudinal/transversal wave speed of the $i$ layer and $\rho_i$ the density of the $i$ layer. In the current analysis the material properties are $E_A=7.31\times 10^{10}$ Pa, $\nu_A=0.325$, $\rho_A=2770$ kg/m$^3$ and $E_B=9.2\times 10^{10}$ Pa, $\nu_B=0.33$, $\rho_B=8270$ kg/m$^3$ for aluminium and brass respectively. Figure \ref{fig:bilayer_sol} shows the comparison between the results obtained with classical FEM (with consistent and lumped mass matrix) SFEM and the analytical solution. Notice that the same number of modes (5 modes), were found with the classical $8\times 8$ finite element and with the lower order $4\times 4$ spectral element. Furthermore, the spectral element, as shown in Figure \ref{fig:bilayer_sol2}, reproduces exactly up to 11 modes, once again showing the improved convergency properties of the spectral method. It is evident that in the considered cases this approximation does not introduce considerable error in the dispersive properties as compared with the full consistent matrix.

\subsection{Matrix with pore/elastic inclusion}
\begin{figure}[H]
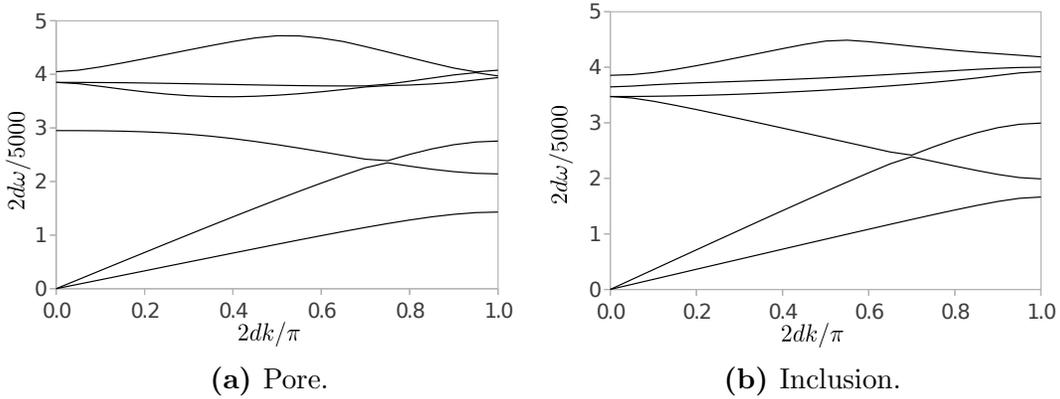

\centering
	\begin{subfigure}[b]{0.45\textwidth}
		\includegraphics[width=\textwidth]{sqpore-8x8-SEM.pdf}
		\caption{Pore.}
	\end{subfigure}\,
	\begin{subfigure}[b]{0.45\textwidth}
		\includegraphics[width=\textwidth]{sqinc-8x8-SEM.pdf}
		\caption{Inclusion.}
	\end{subfigure} 
\caption{Dispersion relations for a square brass inclusion and pore in a square cell of aluminium. Both were computed with $8 \times 8$ SFEM elements.}
\label{fig:pore-inc}
\end{figure}

As a final case we considered the cell of an aluminium matrix with a square pore and a square brass inclusion. Figure \ref{fig:pore-inc} shows the dispersion curves with the $8\times 8$ spectral element used to capture the 11 modes in the previous case. This result is also reported in \cite{Langlet1993}. There is a bandgap with a cut-off frequency close to 3.0 in the cell with the pore. The bandgap however, completely closes in the case of the inclusion since it presents a smaller impedance contrast.
\section{Conclusions}
We have implemented a higher order classical finite element method and spectral finite element method to find the dispersion relations for in-plane $P$ and $SV$ waves, propagating in periodic materials, i.e, phononic crystals. The implemented solution method is based on the Floquet-Bloch theorem from solid state physics, which allows the determination of the dispersion curve from the analysis of a single cell. The numerical dispersion curves of the classical and spectral approaches were used to qualify the performance of each technique in the simulation of mechanical wave propagation problems. Since the dispersion curve simultaneously describes properties of the propagation medium and the external excitation, it is an engineering and objective measure to assess the performance of the numerical method and particularly of its ability to correctly capture the dispersive properties of different material microstructures.

The considered materials correspond to a homogeneous non-dispersive material, a dispersive bilayer material and a composite material exhibiting bandgaps. The composite material is formed by a matrix with a square pore/inclusion. In general, it was found that the spectral finite element method yields reliable results at much higher frequencies than the classical technique, even at the same computational costs. At low frequencies however, the performance of both methods is equivalent and the analysis could rely on the classical algorithm. One of the main advantages of the SFEM is the fact that having used Gauss-Lobatto-Legendre nodes the resulting mass matrix is diagonal by formulation. In the case of an analysis using Bloch-periodic boundary conditions, a diagonal mass matrix allows reducing the generalized eigenvalue problem into a standard eigenvalue problem. This can be accomplished via Cholesky decomposition. 

As a secondary goal, we also evaluated in this work the effects of enforcing diagonality of the mass matrix resulting from a classical algorithm. The dispersion curves obtained with the classical method using a consistent and a lumped mass matrix were the same in practical terms which allows to use such approximation with confidence.

\section{Acknowledgements}
This work was supported by Colciencias and Universidad EAFIT under the Young Researcher and Innovators Program ``Virginia Guti\'errez de Pineda''. The authors would like to thank Professor Anne Christine Hladky-Hennion from  \emph{Institut d'\'electronique de micro\'electronique et de nanotechnologie -Universit\'e Lille Nord}, France for answering so kindly our conceptual queries and also providing us a copy of \cite{Langlet1993}.
\bibliographystyle{gji}
\bibliography{references}
\end{document}